\documentstyle[12pt,fullpage]{article}

\begin{document}
\begin{center}
{\large\bf Magnetic Penetration Depth in a Superconductor}\\
\vskip1cm

 Sang Boo Nam\\
{\sl University Research Center, Wright State University \\
7735 Peters Pike, Dayton, OH 45414 USA}\\
and\\
{\sl Department of Physics, Pohang University of Science and Technology}\\
{\sl Pohang, Kyungbook 790-784, KOREA$^*$}
\vskip2cm

{\bf abstract}
\end{center}

\noindent
It is shown that the notion of the finite pairing interaction energy range $T_d$ results in a linear temperature dependence of the London magnetic penetration depth $\Delta \lambda / \lambda (0) = (T/T_d)(2/\pi)\ln 2$ at low temperature, in the case of the s-wave pairing state.
\vspace*{2cm}

\noindent
PACS numbers 74.25.Ha, 74.20. Fg\\
$^*$e-mail : sbnam@galaxy.postech.ac.kr
\newpage

One of important parameters in a superconductor is the London magnetic penetration depth $\lambda(T)$ which reflects the condensed carrier density, superfluid density $\rho_s(T) / \rho_s (0) = [\lambda(0) /\lambda(T)]^2$.
The temperature dependence of $\rho_s(T)$ plays an important role for understanding the nature of condensation.
In the two fluid picture, $\rho_s(T)$ varies as $1-(T/T_c)^4$.
But the BCS-$\rho_s(T)$ has an activation form at low temperature via the order parameter $\Delta$, which indicates the excitation energy gap.

The measurements of $\lambda(T)$ at low temperature in high $T_c$ superconductors (HTS) are compatible with neither the BCS result nor the two fluid picture. The data of $\lambda(T)$ in a single crystal YBCO [1] indicates a linear temperature dependence.
On the other hand, the data in films [2] behaves the T-square dependence. 
The linear temperature dependence of $\lambda (T)$ is in fact taken as providing evidence that the order parameter has nodes, suggesting the d-wave pairing states [3].
The T-square behavior of $\lambda (T)$ is understood due to scatterings via impurities or defects [4].
Recently, the non-local  effect in a pure d-wave superconductor [5] is suggested to have the T-square behavior of $\lambda (T)$ at very low temperature.  

Contrary to the general belief, I show  here that it is not necessary to have a node in the order parameter, to account for a linear temperature dependence of $\rho_s(T)$ at low temperature.

The fact is that to have a finite value of $T_c$, a finite pairing interaction energy range $T_d$ is required within the pairing theory [6].
In other words, the order parameter $\Delta(k, \omega)$ may be given as
\begin{eqnarray}
\Delta(k, \omega)=\left\{\begin{array}{l} \Delta {\rm~~for}~~|\epsilon_k|< T_d\\
0 {\rm~~~for}~~|\epsilon_k|> T_d\end{array}\right\}
\end{eqnarray}
for all frequencies $\omega$.
Here $\epsilon_k$ is the usual normal state excitation energy with the momentum $k$, measured with respect to the Fermi level.
Here the units of $\hbar = c = k_B = 1$ are used.

By carrying out the $\epsilon_k$-integration of the imaginary part of the usual Green's function [7] consistent with Eq.(1), the density of states $n(\omega)=N(\omega)/N(0)$ is obtained as [6]
\begin{eqnarray}
n(\omega) &=& q(\omega/T_d)+n_{\rm BCS}(\omega)r(\omega/T_d)\\
q(\omega/T_d) &=& (2/\pi)\tan^{-1}(\omega/T_d)\\
r(\omega/T_d) &=& (2/\pi)\tan^{-1}[n_{\rm BCS}(\omega)T_d /\omega]\\
n_{\rm BCS}(\omega) &=& Re \{\omega/(\omega^2-\Delta^2)^{1/2}\}.
\end{eqnarray}
The $q(\omega/T_d)$ is resulted from the $\epsilon_k$-integration of the Green's function with $\Delta(k, \omega)=0$ for $|\epsilon_k| > T_d$.
In the infinite $T_d$ limit, $n(\omega)=n_{\rm BCS}(0)$ as expected. 

In the spirit of the BCS, the London penetration depth $\Delta\lambda = \lambda(T)-\lambda(0)$ may be given as [3]
\begin{eqnarray}
\Delta \lambda/\lambda(0) = \int^\infty_0 d(\omega/T)~n(\omega)f(\omega/T)[1-f(\omega/T)]
\end{eqnarray}
where $f(x) = 1/[1+\exp(x)]$ is the Fermi function.
For the BCS density of states $n_{\rm BCS}(\omega)$, the well known result at low temperature follows
\begin{eqnarray}
[\Delta\lambda/\lambda(0)]_{\rm BCS} = (2\pi \Delta/T)^{1/2}\exp[-\Delta/T].
\end{eqnarray}
By inserting $q(\omega/T_d)$ of Eq.(3) into Eq.(6), at low temperature we get
\begin{eqnarray}
[\Delta\lambda/\lambda(0)]_q = (T/T_d)(2/\pi)\ln 2, 
\end{eqnarray}
similar to that resulted from the order parameter of the d-wave symmetry [3],
\begin{eqnarray}
[\Delta \lambda/\lambda(0)]_d = (T/\Delta_0) \ln 2
\end{eqnarray}
via $n_d(\omega) = \omega/\Delta_0$, where $\Delta_0$ is the maximum value (anti-node) of the order parameter.
The Eq.(8) is a reflection of the finite pairing interaction energy range $T_d$, that is, $q(\omega/T_d)$.
In low $T_c$ superconductor (LTS), the pairing interaction energy range $T_d \simeq T_c \exp (1/g)$ in the weak coupling $g$ limit is large compared to $T_c$ and makes the linear temperature dependence 
of $\lambda(T)$ hardly observable.
On the other hand, in HTS,  even though the exact nature of the pairing interaction is not known, $T_d$ appears to be of the order of $T_c$.
Thus, the linear temperature dependence of $\lambda(T)$ is observed [1].
In fact, the data of [1] yields $T_d \simeq 2T_c$ via Eq.(8).

In conclusion, the notion of the finite pairing interaction  energy 
range results in a linear temperature dependence of superfluid 
density $\rho_s(T)$ at low temperature, which yields 
the T-power laws in the various properties in a superconductor [6].
A linear temperature dependence of $\lambda(T)$ does not imply nodes 
in the order parameter.

I am grateful to members at POSTECH for the warmest hospitality,
and thank members at Physics Department and Institute for Basic Sciences
at POSTECH, and the Center for Theoretical Physics at Seoul
National University for giving me the opportunity to have discussions 
with my old teachers and old friends as well as new friends.

\newpage
{\large\bf References}
\vskip1cm

\begin{itemize}
\item[[1]] W. N. Hardy et al, Phys. Rev. Lett. \underline{70}, 3999 (1993).
\item[[2]] Z. Ma et al, Phys. Rev. Lett. \underline{71}, 781 (1993).
\item[[3]] D. J. Scalapino, Phys. Rep. \underline{250}, 329 (1995).
\item[[4]] P. J. Hirschfeld and N. Goldenfeld, Phys. Rev. \underline{B48}, 4219 (1993).
\item[[5]] I. Kosztin and A. J. Leggett, Phys. Rev. Lett. \underline{79}, 135 (1997).
\item[[6]] S. B. Nam, Phys. Lett. \underline{A193}, 111 (1994), ibid(E) \underline{A197}, 458 (1995).
\item[[7]] J. R. Schrieffer, Theory of Superconductivity (Benjamin, New York, 1964) ch. 7.
\end{itemize}
\end{document}